\documentclass[10pt]{article}
\usepackage{epsf}

\newlength{\dinwidth}
\newlength{\dinmargin}
\def\lapproxeq{\lower .7ex\hbox{$\;\stackrel{\textstyle <}{\sim}\;$}}
\def\gapproxeq{\lower .7ex\hbox{$\;\stackrel{\textstyle >}{\sim}\;$}}

\def\be{\begin{equation}}
\def\ee{\end{equation}}
\def\bea{\begin{eqnarray}}
\def\eea{\end{eqnarray}}

\def\bb{{b\bar{b}}}

\begin{document}
\titlepage

\begin{flushright}
IPPP/04/19 \\
DCPT/04/38\\
3rd June 2004 \\
\end{flushright}

\vspace*{4cm}

\begin{center}
{\Large \bf Can an invisible Higgs boson be seen via diffraction at the LHC?}

\vspace*{1cm} \textsc{K. Belotsky$^{a,b}$, V.A.~Khoze$^{c,d}$, A.D. Martin$^c$ and M.G. Ryskin$^{c,d}$} \\

\vspace*{0.5cm} $^a$ Moscow Engineering Physics Institute,
Moscow, Russia \\[0.5ex]
$^b$ Center for Cosmoparticle Physics "Cosmion" of
Keldysh Institute of Applied Mathematics, \\
Moscow, 125047, Russia \\[0.5ex]
$^c$ Department of Physics and Institute for
Particle Physics Phenomenology, \\
University of Durham, DH1 3LE, UK \\[0.5ex]
$^d$ Petersburg Nuclear Physics Institute, Gatchina,
St.~Petersburg, 188300, Russia \\

\end{center}

\vspace*{1cm}

\begin{abstract}
We study the possibility of observing an `invisible' Higgs boson in central
exclusive diffractive production at the LHC.  We evaluate the cross section using,
as a simple example, the Standard Model with a heavy fourth generation, where
the invisible decay mode $H \to \nu_4 \bar{\nu}_4$ dominates, with the heavy
neutrino mass $M(\nu_4) \simeq 50$ GeV.  We discuss the possible
requirements on trigger
conditions and the  background processes.
\end{abstract}


\section{Introduction}

There exist several extensions of the Standard Model (SM) in which
the Higgs boson decays dominantly into particles
which cannot be directly detected. One example is the Standard Model
with a fourth generation\cite{4gen,nu4,Shibaev}, where the invisible decay mode $H \to \nu_4 \bar{\nu}_4$
may occur with a large branching fraction\cite{nu4}.
 In the case of supersymmetry, the Higgs can decay with a
large branching ratio into
gravitinos or neutralinos or other neutral supersymmetric particles, see, for example \cite{HDDT}.
Yet another possibility are models with large extra dimensions, see, for example, Refs.~\cite{ADD,RS,GRW,huitu}

Searches for an invisible Higgs at the LHC have been addressed recently in Refs.~\cite{Niki,MB,Godbole, Kersevan}.
One proposal is to observe such an `invisible' Higgs in inelastic events with large missing
transverse energy, $/\!\!\!\! E_T$, and two high $E_T$ jets\cite{Zepp,Niki}. Here the Higgs boson is
produced by $WW$ fusion and therefore has large transverse momentum.  However, in such a
process, it is not possible to measure the mass of the boson.  Moreover even the
quantum numbers are not known.  Other undetected neutral
particles (for example, neutrinos, photinos) may be produced which
carry away large $/\!\!\!\! E_T$. From this viewpoint central exclusive
diffractive production, $pp \to p+H+p$ looks to be a much
more favourable process, see, for example, Ref.~\cite{AR,KMRmm}.  First, the mass $M_H$ can be accurately determined by
observing the forward going protons and measuring the missing mass\cite{AR,DKMOR}. The existence of
the sharp peak in the missing mass spectrum dramatically reduces any background contributions.
Second,
we have information about the quantum numbers of the object produced by Pomeron-Pomeron fusion:
the boson must be neutral, colourless, flavourless and have natural parity, $P=(-1)^J$, with
$J^P=0^+$ being by far the most likely\cite{KMRProsp,KMRcpx}\footnote{The spin-parity
may be studied in more detail by measuring the angular correlation between the transverse
momenta of the outgoing protons\cite{KKMRCentr}.}.

Moreover, in some popular models (in particular, the minimal supersymmetric model (MSSM)
with large values of tan$\beta$\cite{CH}) the coupling of the Higgs to $W,Z$ bosons is suppressed
while the coupling to gluons is enhanced\cite{BDMV}.   Thus the inelastic production cross section
with large missing $E_T$ is suppressed, while the exclusive signal is enhanced by about an
order-of-magnitude as compared to the SM prediction\cite{KKMRext}.

\section{A typical cross section}

To illustrate the possibility of using central exclusive diffractive production,
$pp \to p+H+p$, to observe an invisible Higgs $H$ we take a simple example.
We consider the SM with an additional fourth generation of heavy
fermions
\footnote
{There are arguments, both from cosmology and astrophysics\cite{4gen, 4gen-a}
and from precision fits of
electroweak data\cite{Okun}, in favour of the presence of a heavy fourth generation.}.
The mass of the heavy neutrino must be greater than $M_Z/2$, since it has not been
seen in $Z$-decays at LEP\footnote{A recent detailed analysis of the $Z$-resonance shape
shows that the fourth generation is excluded at 95\% C.L. for
$M(\nu_4)<46.7 \pm 0.2$ GeV\cite{okun1}.}. On the other hand cosmological constraints imply
that it cannot be too heavy; a mass greater a few TeV is likely to be excluded.
Data on direct searches for Weakly Interacting Massive Particles (WIMP) in underground installations
suggest that 4th neutrino mass is not much above 50 GeV \cite{4gen-a}.

We therefore
take $M(\nu _4)=50$ GeV for our numerical estimates, and assume that
all the other (charged) fermions of the fourth generation
have masses above the experimental lower limits; to be specific we assume that they
are heavier than the
top quark.

In such a scenario the branching fractions of the Higgs boson into the usual
visible decay channels are suppressed due to the large $H \to \nu_4 \bar{\nu}_4$
decay width.  In particular, for a Higgs of mass 120 GeV, the
$H \to b \bar{b}$ decay
is reduced from the SM fraction of 68\% to about 10\% if there is a heavy fourth
generation\footnote{Note that for a wide range of Higgs masses, the existence of a fourth
generation would induce a significant reduction of the two-photon decay width of
the Higgs, see, for example, \cite{GIS,nu4}.}.
On the other hand the $gg \to H$ coupling becomes about three times larger, as we now have
three types of heavy quarks
($U$ and $D$ quarks of the fourth generation and the top quark)
with $m_Q>M_H/2$.  Thus the central exclusive Higgs production
cross section is enhanced by up to a factor of 9 from the SM expectation, see, for example, Ref.~\cite{GIS,nu4}.
To calculate the cross sections for a range of Higgs masses, we use the
method described in Refs.~\cite{KMRProsp,KKMRCentr,KMRsoft,KMRmm}.  The results, at the LHC
energy, are presented in Table 1.  We show the total exclusive cross section,
$\sigma~=~\sigma(pp \to p+H+p)$, and the visible component, $\sigma(\bb)$, going through the $\bb$ decay.

\begin{table}[htb]
$$\begin{array}[t]{|l c c c c|} \hline
M_{\rm H}~{\rm (GeV)}  & 120 & 150 & 180 & 210
\rule[-1.5ex]{0ex}{4.5ex}      
\\ \hline
\Gamma(H\to gg)~{\rm (MeV)}  & 2.2 & 4.1 & 6.9 & 10.8 \\
\sigma~{\rm (fb)} & 21 & 11 & 5.9 & 3.6 \\
{\rm Br}(\bb)  & 11\% & 4\% & 0.5\% & 0.2\% \\
\sigma(\bb)~{\rm (fb)}  & 2.3 & 0.4 & 0.03 & 0.007
\\ \hline
\end{array}$$
\caption{The total cross section, $\sigma$, for the exclusive double-diffractive production
of a Higgs boson of mass $M_H$ at the LHC, assuming that a fourth heavy generation of fermions
exists.  Most of the cross section corresponds to the invisible decay $H \to \nu_4 \bar{\nu}_4$.
We take $M(\nu_4)$ to be 50 GeV.
The component $\sigma(\bb)$ corresponding to the visible $H \to \bb$ decay
is also given.  The $H\to gg$ and $H\to bb$ decay widths and branchings have been calculated
using the HDECAY code\cite{HDECAY}.}
\end{table}

Interestingly, for a Higgs mass of 120 GeV, $\sigma(\bb)$ remains close to the
conventional SM prediction, even though the branching fraction is reduced by a factor of about 7.
The suppression is compensated by the enhancement of the $H \to gg$ width by a factor of up to 9.
Using the cuts and efficiences of Ref.~\cite{DKMOR}, we see, from the estimates
given in that paper, that we expect about 10
$p+H(\to \bb)+p$ events with a QCD $\bb$ background of 4
events\footnote{Note  that in \cite{DKMOR} the quantity $\Delta M_{\rm miss}, $
which was called the mass resolution, is not the dispersion, $\sigma_{\rm mass}$,
of the Gaussian distribution over the missing mass, but rather the
integration range for the signal.  In order to detect ~95\% of the signal
$\Delta M_{\rm miss}$ should be equal to $4\sigma_{\rm mass}$. Therefore,  if
$\sigma_{\rm mass}$ were equal to 1 GeV, then the ratio of the signal
will decrease by a factor of about 4.},
for a luminosity
${\cal L} = 30~{\rm fb}^{-1}$.

However for higher Higgs masses ($M_H>150$ GeV), the cross section, $\sigma(\bb)$, for the visible
$\bb$ signal appears to be too small to be observable at the LHC, at least
for an integrated lumionsity of the order of $30~{\rm fb^{-1}}$.   On the other hand, the exclusive cross
section in the invisible Higgs mode is still large enough to give a detectable
event rate; even for $M_H=210$ GeV we have a cross section of nearly 4 fb.

\section{Triggering for an invisible Higgs}

At first sight the best signal for an invisible Higgs decay mode
would be an `empty event' trigger with no deposition in the central
detector (CD).
However elastic $pp$ scattering gives a huge number
of such events.  For a luminosity of ${\cal L}=10^{33}~{\rm cm}^{-2}{\rm s}^{-1}$ we would
anticipate more than 20 million background events/sec, since $\sigma_{\rm
el} \geq 20$ mb.
Thus the only way to isolate signal events is to use information from the forward proton detectors,
which register protons with a fraction $x_p=1-\xi$ of their incoming energy.
If events are selected where both forward protons have lost some small fraction,
 $\xi\sim 0.001-0.01$, of their initial energy, then, first, the
elastic interaction is eliminated, and, next, (just by kinematics) rapidity
 gaps ($y_i > \ln 100\sim 4.5$) are generated around the protons.
In this case, the main background is generated by `soft' inelastic
Pomeron-Pomeron fusion.  Up to a factor of two uncertainty, the cross
section for the production of the central system of low $p_t$ particles,
with the invariant mass $M$ in the range of 100-200 GeV, is \cite{KMRsoft,KMRProsp}
\begin{equation}
\frac{d\sigma^{CD}}{dy_1dy_2}~= ~4 ~\mu {\rm b},
\end{equation}
with a weak mass dependence\footnote{Note that this estimate
is in a qualitative agreement with the old results \cite{AKLR} obtained
within both, the triple-Pomeron approach and the one-pion-exchange model}.
Here $y_i = {\rm ln}(1/\xi_i)$ are the rapidity intervals which separate the
centrally produced system from the outgoing protons\footnote{Strictly speaking
the $y_i$ are the rapidity intervals between the respective proton (with energy $E_0$)
and the most energetic hadron in the
central system with the same transverse mass ($m_T=\sqrt {m_{had}^2+p^2_t}=m_p$)
and energy $E=\xi_i E_0$.   For a light hadron at
low $p_t$, the intervals will be smaller by the constant amount ${\rm ln}(m_p/m_T)$}.
The mass and the c.m. rapidity of the centrally
produced system are
\begin{equation}
M^2 \simeq \xi_1 \xi_2 s ~~~{\rm and}~~~y=(y_1-y_2)/2.
\end{equation}
If we integrate over the available rapidity interval,
$-2.5 < y < 2.5$  ($\Delta y \sim 5$),
and account for the mass bin, $M=100-200$ GeV,
 we find
$$\sigma^{CD} ~\sim ~30 ~ \mu {\rm b}.$$
For a luminosity of ${\cal L}=10^{33}~$cm$^{-2}$sec$^{-1}$, we thus have a rate of about 30 KHz.

For the off-line analysis we would need to know only the momenta of the
two tagged protons and that no other secondaries are produced.  This is
rather a limited amount of information and, ideally, we may even
contemplate accumulating the information at 30 KHz.  Moreover, one
could decrease  the rate by installing a veto detector to reject events
containing secondaries in the central rapidity region.  Unfortunately,
in practice, such an ideal experimental set-up is very difficult to
achieve.  The LHC general purpose detectors - CMS and ATLAS - are
designed, and optimised, for `discovery physics' with a high $p_t$
trigger in the central rapidity region.  It is not possible to trigger
on `nothing' with these detectors. In particular, there will always be
noise in the calorimeters.

 On the other hand, for the present purpose, it is not necessary to measure the
momenta of these secondaries.  At the trigger level, it would be
sufficient to supplement the Central Detector with an
additional simple detector just to suppress events in which new
particles are emitted. Indeed, at the trigger level, it would be
enough to detect only charged
(or only neutral, say, photons)
particles in a limited rapidity interval, $|y|=3-6$, not currently
covered  by the  central calorimeter\footnote{Such particles will be missed
 by the central CMS and ATLAS detectors,  but charged particles may  be detected by the
planned TOTEM T1 and T2 detectors \cite{RO2,Totem}.
Moreover, a combination of the signal from the forward proton detectors and the
veto from the T1 and T2 detectors can help to reduce the rate
of background events and to improve the trigger budget. The rate limit
of about 1 KHz could be set within the present Level-1 trigger  budget
for this channel.}.  At the final stage, in the off-line analysis the
information from the Central Detector can be used to suppress the
physical background more effectively.

Can we use information from the proton taggers for triggering?
This issue is under study by the experimental collaborations.
Various LHC beam optics are being considered\cite{RO}, one of which
would require forward proton detectors
(roman pots or microstations) to be installed at about 300 - 400 m
from the interaction point.   Due to their distant position, the signals from
these proton taggers are delayed, and it is difficult to use them
as a Level-1 trigger.  However future electronics may allow an extension of
the trigger latency (decision making time) to 300 - 400 m.
Alternative beam optics, currently under investigation, which would avoid
the above problem, is based on roman pots
at about 200 m or less from the interaction point\cite{RO1}.  So the
situation is not resolved yet.

Note that due to `pile-up' events~--- that is several independent $pp$
interactions in a single bunch crossing~--- it would be better to work
at a relatively low luminosity.
Indeed, we need to collect
only events which have no inelastic interaction in a
bunch crossing.  Thus, assuming that a supplementary
detector is used
as a veto trigger, the effective luminosity appropriate for the Level-1
trigger would be
\begin{equation}
 {\cal L}_{\rm eff}~=~{\cal L}_{0}~{\rm
exp}(-n({\cal L}_0)), \label{eq:A1}
\end{equation}
where $n({\cal L}_0)$ is the
average number of pile-up events at collider luminosity ${\cal L}_0$.
For  ${\cal L}_0=10^{33}~{\rm cm}^{-2}{\rm s}^{-1}$ we expect $n({\cal
L}_0) \simeq 2.3$, which gives
${\cal L}_{\rm eff}=10^{32}~{\rm cm}^{-2}{\rm s}^{-1}$.  As a result even
at the level-one, the expected
detection rate of the events with the rapidity gaps
will be 10 times less,
and the
effective integrated luminosity would be ${\cal L}_{\rm eff} \simeq 3~{\rm
fb}^{-1}$, and not 30 ${\rm fb}^{-1}$ (as it is expected for the run
with ${\cal L}_0=10^{33}~{\rm cm}^{-2}{\rm s}^{-1}$).
  If we would assume that the efficiency of tagging both the
forward protons is $\epsilon ~$=~0.6 \cite{DKMOR}, then about 10 events
should be registered for ${\cal L}_{\rm eff}=3~{\rm fb}^{-1}$ and
$M_H=180$ GeV.
Note that the maximal value of the effective luminosity,
${\cal L}_{\rm eff}=1.5\times 10^{32}~{\rm cm}^{-2}{\rm s}^{-1}$, is reached at a collider luminosity
of ${\cal L}_0=4\times 10^{32}~{\rm cm}^{-2}{\rm s}^{-1}$.\\

\section{Background to the invisible Higgs signal}

The crucial point for detecting an invisible Higgs boson is the size of the background.
The only signal which is detected for
the process is the registration of the two outgoing protons with a fraction $x_p~=~1-\xi$
of the initial proton momentum just less than 1; typically $\xi \simeq$ 0.01.  Unfortunately
an outgoing proton can loose part of its energy simply by QED radiation\cite{KMRmm}.  The probability to emit a
photon of energy $\omega$ is\cite{QED}
\be
\frac{2\alpha}{3\pi} \frac{\langle q^2_T \rangle}{m_p^2} \frac{d\omega}{\omega},
\label{eq:QED}
\ee
where $q_T \ll m_p$ is the transverse momentum of the outgoing proton of mass $m_p$ and $\alpha=1/137$
is the QED coupling.  Thus the cross section for quasielastic scattering, $pp \to (p\gamma)+(p\gamma)$,
which may mimic a missing mass invisible Higgs event, is
\be
\frac{d\sigma}{dy}|_{y=0}~~\sim~~
\left( \frac{2\alpha}{3\pi} \frac{\langle q^2_T \rangle}{m_p^2} \right)^2 ~\frac{\Delta M^2}{M^2}~\sigma_{\rm el}(pp)~~
\simeq~~3~{\rm pb}.
\label{eq:QED2}
\ee
To obtain this numerical estimate we assume an integration range of $\Delta M$ = 1 GeV
(see \cite{DKMOR}) and $M_H$ = 120 GeV.
We have also taken $\sigma_{\rm el}$ = 25 mb and $\langle q^2_T \rangle = 1/B_{\rm el} = 0.05~{\rm GeV}^2$
at the LHC energy\cite{KMRsoft}.
To compare the background with the cross sections presented in Table 1 we have to integrate
(\ref{eq:QED2}) over the rapidity interval in which the Higgs signal is collected.  For instance, for a
Higgs of mass $M_H \sim 150$ GeV this interval is $|y|<2.5$.
The cross section of (\ref{eq:QED2}) therefore has to be integrated over the rapidity interval
$\Delta y \simeq 5$.  Thus we see the QED background cross section $\sigma_{\rm QED} \simeq$ 15 pb
exceeds the invisible Higgs signal, $\sigma \simeq$ 10 fb by almost a factor of 1500. Thus
to suppress the QED background it is necessary to have forward veto electromagnetic calorimeters to detect
the forward photons, with energies about $M_H/2 \simeq ~$75 GeV, with an efficiency better
than about $1-1/{\sqrt{1000}}$, that is 97\%\footnote{Also it is
necessary to allow for the radiative tail which will spread out the shape of the Higgs peak.}.
This may be achieved, at least to some extent, by the CMS detector by using its zero degree
calorimeter (ZDC), which may allow the detection of forward photons with energies above 50 GeV.

Another source of background is double-diffractive dissociation,
$pp \to X+Y$, where both the excited $X$ and $Y$ systems contain a
proton with $\xi \sim 0.01$. The selection  $\xi_{1,2} \sim 0.01$
means that we already have rapidity gaps between the forward
protons and the central system.  The main danger is to have
`quasi-elastic' Pomeron-Pomeron scattering in the central region,
$PP \to h_1 h_2$ where the hadrons $h_i$ may be, for example, the
$f_0 (0^{++})$ or $f_2 (2^{++})$mesons or even a glueball.  Quantum numbers prevent a
$h_i$ being a single pion.  The quasi-elastic process produces low
multiplicity events (typically at least 4 pions) with a third
rapidity gap in the central region, as shown in Fig.1.
\begin{figure}
\begin{center}
\centerline{\epsfxsize=7cm\epsfbox{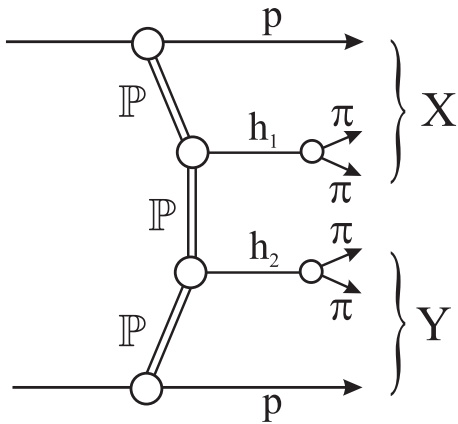}} \caption{A schematic diagram of the double-diffractive dissociation
background, $pp \to X+Y$, to an invisible Higgs signal
\label{fig}}
\end{center}
\end{figure}
To estimate this background we use the model of Ref.~\cite{KMRsoft}.
The cross section of the process shown in Fig.1 is given by
\be
\frac{d\sigma^{DD}}{dy_1dy_2}~\sim~
{\cal L}_{PP}~\frac{\sigma_{PP}^2}{32\pi^3 B_{PP}}~= ~1 - 100 ~{\rm nb},
\label{eq:DD1}
\ee
where the effective Pomeron-Pomeron luminosity is\cite{KMRsoft}
$${\cal L}_{PP}~\sim ~0.4 \times 10^{-3}, $$
the Pomeron-Pomeron cross section in
a low mass (resonance) region is $\sigma_{PP}=1 - 10$ mb, and the
`elastic' Pomeron-Pomeron scattering slope is
$B_{PP}\sim 1 - 2 ~{\rm GeV}^{-2}$.
When we integrate over the available rapidity interval of the central system, $\Delta y \sim 5$,
and account for the mass resolution,
$$\Delta M^2/M^2 ~=~2\Delta M/M ~\sim~ 0.02,$$
 we find a background of
$$\sigma^{DD} ~\sim ~0.1 - 10 ~{\rm nb}.$$
An even more
conservative\footnote{The analysis of Ref.~\cite{UA8} does not account
for the gap survival probability, ${\hat S}^2$. Therefore the value
$\sigma^{\rm eff}_{PP}=1.5$ mb claimed by UA8 should be considered as
the product ${\hat S}^2\sigma_{PP}$ corresponding to a `corrected' $\sigma_{PP}
\sim 10 - 20$ mb. Also note that the UA8 evaluation\cite{UA8}, $\sigma^{DD} \sim 300~$ nb,
does not take into account the decrease in the survival
probability of the rapidity gaps of the three-gap process
(by a factor 2-3) as we go from the energy, $\sqrt {s} = 630$ GeV, of the UA8 experiment
up to the LHC.} evaluation, based on the effective Pomeron-Pomeron cross section $\sigma_{PP} \simeq 1.5~$mb
measured by the UA8 collaboration\cite{UA8}, gives $\sigma^{DD} \sim 300~$nb.  The centrally produced
$h_1 h_2$ system will decay into at least 4 pions.   Thus we need to be
sure that at least one of the four pions
 will be observed.
That is the probability to detect one pion must be
better than
\be
1-\left( \frac{S}{B} \right)^{\frac{1}{4}}~=~1~-~\left(\frac{\sigma(pp \to p+H+p)}{\sigma^{DD}} \right)^{\frac{1}{4}}
~\sim~1~-~\left(\frac{10 ~{\rm fb}}{300~{\rm nb}} \right)^{\frac{1}{4}}~\sim~ 99\%,
\label{eq:pion}
\ee
where $S/B$ is the ratio of the invisible Higgs signal to the double-diffractive
dissociation background.
Since, here, we have used $\sigma^{DD}$=300 nb, it is a very conservative estimate.  Even so, it appears to
be a realistic requirement for the detection of the background. Of course the suppression of the
background will depend
on the coverage of the detectors.  In this connection, we note, that in the `worst' possible decay
configuration, at least one
pion must have rapidity $|y_\pi|<5$, simply from kinematic considerations
\footnote{The typical energy of the two fast pions is about
$(M_H/4)e^{y_H}$, where $M_H$ and $y_H$ denote the mass and rapidity of
the central system measured by the forward detectors (see (2)).
For the two slow pions the energies are around $(M_H/4)e^{-y_H}$.
The expected c.m. rapidities of these four pions are in the region
of $|y_\pi|\sim 4 - 8$.}.
Of course, a detailed evaluation of the
size of the background will require a Monte Carlo simulation of the response of the available detectors to
the double-diffractive dissociation events.

\section{Summary}

We have shown that there is a good chance to observe a Higgs boson which decays invisibly
via central exclusive diffractive production, $pp \to p+H+p$.
Contrary to conventional
inelastic production, the mass of the Higgs boson can be accurately measured by the
missing-mass method. This is a crucial ingredient in reducing the background to
the level of the signal.  Moreover for exclusive process, it is known that the produced
object is flavourless and is a colour singlet.  Due to the `pile-up' problem, it will be
most effective to work at luminosities in the
range ${\cal L}_{\rm eff}=10^{32}-10^{33}~{\rm cm}^{-2}{\rm s}^{-1}$.

To suppress the background arising from QED radiation and/or soft double-diffractive
dissociation, the Central Detector should be supplemented with forward calorimeters
able to reject events with additional high energy photons and charged pions with
energies in the range 5-200 GeV.  In order to reach a signal-to-background ratio
$S/B>1$, these detectors should have an efficiency of photon or pion registration
of typically 98\%.

\section*{Acknowledgements}

We are grateful M. Yu. Khlopov for initiating this paper. We thank
M. Boonekamp, D. Fargion, R.V. Konoplich, K. Osterberg and S. Tapprogge
and, especially A. De Roeck, A. Nikitenko and R. Orava,
for useful discussions.
ADM thanks the Leverhulme trust for an Emeritus Fellowship and MGR thanks the IPPP at the University of
Durham for hospitality. This work was supported by a Royal Society special project grant, by
the UK Particle Physics and Astronomy Research Council, by grants RFBR 04-02-16073; RFBR 02-02-17490; UR.02.01.008
and by the Federal Program of the Russian Ministry of Industry, Science and Technology
SS-1124.2003.2; 40.022.1.1.1106.

\end{document}